\documentclass[preprint,eqsecnum,aps,nofootinbib]{revtex4}
\usepackage{amsfonts,amsmath,amssymb,amsthm}
\usepackage{latexsym}
\usepackage{bbm,bm}
\usepackage{graphicx}
\usepackage{tikz} 

\newcommand{\beq}{\begin{equation}}
\newcommand{\eeq}{\end{equation}}
\newcommand{\beqs}{\begin{eqnarray}}
\newcommand{\eeqs}{\end{eqnarray}}

\begin{document}

\title{ R\'{e}nyi and von Neumann entropies of thermal state in Generalized Uncertainty Principle-corrected harmonic oscillator}

\author{MuSeong Kim$^1$, Mi-Ra Hwang$^1$, Eylee Jung$^1$, and DaeKil Park$^{1,2}$\footnote{dkpark@kyungnam.ac.kr} }

\affiliation{$^1$Department of Electronic Engineering, Kyungnam University, Changwon
                 631-701, Korea    \\
             $^2$Department of Physics, Kyungnam University, Changwon
                  631-701, Korea    
                      }

\begin{abstract}
The  R\'{e}nyi and von Neumann entropies of thermal state in generalized uncertainty principle (GUP)-corrected single harmonic oscillator system are explicitly computed within the first order of  
GUP parameter $\alpha$. While the von Neumann entropy with $\alpha = 0$ exhibits a monotonically increasing behavior in external temperature, the nonzero GUP parameter makes a decreasing behavior 
at large temperature region. As a result, for the case of $\alpha \neq 0$, the von Neumann entropy is maximized at the finite temperature $T_*$. The R\'{e}nyi entropy $S_{\gamma}$ with nonzero $\alpha$ also exhibits similar behavior 
at large temperature region. In this region the R\'{e}nyi entropy exhibits a decreasing behavior with increasing temperature. The decreasing rate becomes larger when the order of the R\'{e}nyi entropy is smaller. 
\end{abstract}

\maketitle

\section{Introduction}

We believe that at the Planck scale usual quantum mechanics and relativity are not expected to apply, and the effects of quantum gravity are expected to dominate. Although we do not understand the quantum gravity fully, it is also believed 
that there exists a minimal length (ML) at this scale. 
The existence of the ML at the Planck scale seems to be a universal characteristic of quantum gravity\cite{townsend76,amati89,garay94}. It appears in loop quantum gravity\cite{rovelli88,rovelli90,rovelli98,carlip01},
string theory\cite{konishi90,kato90,strominger91}, path-integral quantum gravity\cite{padmanabhan85,padmanabhan85-2,padmanabhan86,padmanabhan87,greensite91}, and black hole physics\cite{maggiore93}. 
Such a ML also appears in some microscope thought-experiment\cite{mead64}. 

From an aspect of quantum mechanics the existence of ML modifies the uncertainty principle from Heisenberg uncertainty principle (HUP)\cite{uncertainty,robertson1929}
$\Delta P \Delta Q \geq \frac{\hbar}{2}$ 
to generalized uncertainty principle (GUP)\cite{kempf93,kempf94}. This is because of the fact that the uncertainty of the position $\Delta Q$ should be larger than the ML. In order to examine the effect of the modified quantum mechanics the ML and GUP are recently used to explore 
several branches of physics such as quantum mechanics\cite{chung19}, quantum electrodynamics\cite{bosso18}, quantum cosmology\cite{bosso19}, quantum gravity\cite{modak19}, and black hole physics\cite{xiang18}. 
More recently, the Feynman propagators\cite{feynman,kleinert,jizba12} for the GUP-corrected quantum mechanics were derived in $1d$ free particle system \cite{das2012,gangop2019}, $1d$ harmonic oscillator\cite{comment-1}, and 
$d$-dimensional harmonic oscillator\cite{park-20}. Moreover, the singular potential problem in the GUP-corrected quantum mechanics was discussed in Ref. \cite{point-1}. 

Quantum information theory (QIT)\cite{text} has been rapidly developed recently due to their applications to various quantum information
processing such as quantum teleportation\cite{teleportation},
superdense coding\cite{superdense}, quantum cloning\cite{clon}, quantum cryptography\cite{cryptography,cryptography2}, quantum
metrology\cite{metro17}, and quantum computer\cite{qcreview,computer}.  
In QIT the von Neumann entropy\cite{text} of quantum state $\rho$ measures the degree of information contained in $\rho$. It is a direct counterpart to the 
Shannon entropy in classical information theory. There are several other entropic quantities introduced in the information theories. In particular, R\'{e}nyi entropies\cite{renyi60,Bengtsson} of order $2$ have many 
interesting properties such as the strong subadditivity inequality\cite{adesso12} for tripartite quantum states. R\'{e}nyi entropies have been applied in the study of quantum metrology\cite{metro}, entanglement measure\cite{emeasure}, channel capacities \cite{renyicapacity}, work value of information \cite{workvalue}, and entanglement spectra in many-body systems \cite{renyispectrum}. Experimental accessibility of R\'{e}nyi entropies is also discussed in specific quantum protocols\cite{protocol}.

The well-known example of the Planck scale is an early universe ($t \leq 10^{-43} (s)$ after big bang). However, we do not know how QIT plays any role at this early stage. Of course, study on the role of QIT at the Planck scale may need 
long-term research. As a first step toward this issue, we will examine in this paper how several quantum information quantities are modified due to the presence of ML by making use of the simple GUP.
In order to explore this issue we derive the thermal state $\rho_T$ of the simple harmonic oscillator by making use of 
Ref. \cite{comment-1}. \textcolor[rgb]{0.00,0.00,1.00}{Then, we will compute several QIT quantities of $\rho_T$ such as purity function , R\'{e}nyi entropy, and von Neumann entropy explicitly.
Finally, we will examine how those quantities are modified due to GUP or ML. }


The specific form of GUP we will choose is  \textcolor[rgb]{1.00,0.00,0.00}{a form}  proposed in Ref. \cite{kempf94}:
\begin{equation}
\label{GUP-d-1}
\Delta P_i \Delta Q_i \geq \frac{\hbar}{2} \left[ 1 + \alpha \left\{ (\Delta {\bf P})^2 + \langle \widehat{{\bf P}} \rangle^2 \right\} + 2 \alpha \left\{ (\Delta P_i)^2 + \langle \widehat{P}_i \rangle^2 \right\} \right]
\hspace{1.0cm}  (i = 1, 2, \cdots, D)
\end{equation}
where $(\Delta {\bf P})^2 = \sum_{j=1}^D (\langle \widehat{P}_j^2 \rangle - \langle \widehat{P}_j \rangle^2 )$, $\langle \widehat{{\bf P}} \rangle^2 = \sum_{j=1}^D \langle \widehat{P}_j \rangle^2$,  and $\alpha$ is a GUP parameter, which has a dimension $(\mbox{momentum})^{-2}$. Using 
$\Delta A \Delta B \geq \frac{1}{2} | \langle [\widehat{A}, \widehat{B}] \rangle |$, Eq. (\ref{GUP-d-1}) induces the 
modification of the commutation relation as\footnote{\textcolor[rgb]{1.00,0.00,0.00}{ In fact, if we choose $\left[\widehat{P}_i, \widehat{P}_j \right] = 0$, the Jacobi identity determines 
$\left[ \widehat{Q}_i, \widehat{Q}_j \right]$ in a form
$$\left[ \widehat{Q}_i, \widehat{Q}_j \right] = \frac{4 \alpha^2 \widehat{{\bf P}}^2}{1 + \alpha \widehat{{\bf P}}^2} \left( \widehat{P}_i \widehat{Q}_j - \widehat{P}_j \widehat{Q}_i \right) = {\cal O} (\alpha^2).$$
Since our discussion in this paper will be only up to first order of $\alpha$, we can simply assume $ \left[ \widehat{Q}_i, \widehat{Q}_j \right] = 0$.}}
\begin{eqnarray}
\label{GUP-d-2}
&& \left[ \widehat{Q}_i, \widehat{P}_j \right] = i \hbar \left( \delta_{ij} + \alpha \delta_{ij} \widehat{{\bf P}}^2 + 2 \alpha \widehat{P}_i \widehat{P}_j  \right)    \\    \nonumber
&& \hspace{1.0cm} \left[ \widehat{Q}_i, \widehat{Q}_j \right] = \left[\widehat{P}_i, \widehat{P}_j \right] = 0.
\end{eqnarray}

The existence of the ML is easily shown at $d = 1$. In this case  Eq. (\ref{GUP-d-1}) is expressed as 
\begin{equation}
\label{GUP-d-3}
\Delta P \Delta Q \geq \frac{\hbar}{2} \left\{ 1 + 3 \alpha (\Delta P)^2 \right\}
\end{equation}
if $\langle \widehat{P} \rangle = 0$. Then\footnote{One can consider more general one-dimensional GUP $$\Delta P \Delta Q \geq \frac{\hbar}{2} \left\{ 1 + \alpha (\Delta Q)^2 + \beta (\Delta P)^2 + \gamma \right\},$$ which leads both 
ML and minimal momentum. However, it seems to be highly difficult to derive the thermal state with this general GUP. Hence, we will use the simple GUP (\ref{GUP-d-3}) in this paper.}  , the equality of Eq. (\ref{GUP-d-3}) yields 
\begin{equation}
\label{minimal-length}
\Delta Q^2 \geq \Delta Q_{min}^2 = 3 \alpha \hbar^2.
\end{equation}
If $\langle \widehat{P} \rangle \neq 0$, the corresponding ML becomes $\Delta Q_{min}^2 = 3 \alpha \hbar^2 \left[ 1 + 3 \alpha \langle \widehat{P} \rangle^2 \right]$, which is different from Eq. (\ref{minimal-length}) at ${\cal O} (\alpha^2)$. 
%


If $\alpha$ is small, Eq. (\ref{GUP-d-2}) can be solved as 
\begin{equation}
\label{GUP-d-4}
\widehat{P}_i = \widehat{p}_i \left(1 + \alpha \widehat{{\bf p}}^2 \right) + {\cal O} (\alpha^2)   \hspace{1.0cm} \widehat{Q}_i = \widehat{q}_i
\end{equation}
where $p_i$ and $q_i$ obey the usual Heisenberg algebra $[q_i, p_j] = i \hbar \delta_{ij}$. We will use Eq. (\ref{GUP-d-4}) in the following to compute the R\'{e}nyi and von Neumann entropies within the 
first order of $\alpha$. 

The paper is organized as follows. In next section we will derive the thermal state $\rho_T$ of the GUP-corrected harmonic oscillator system. 
Since the thermal state is not Gaussian when $\alpha \neq 0$, it is highly difficult to solve the eigenvalue equation of $\rho_T$. 
However, it is shown in the following sections that the eigenvalue can be derived up to the first order of $\alpha$ without solving the eigenvalue 
equation explicitly. In section III we compute the ${\mbox Tr} \rho_T^n$ explicitly within ${\cal O} (\alpha)$. The purity function is also explicitly 
derived. It is shown in this section that while the thermal state with $\alpha = 0$ transforms from pure to completely mixed state with increasing the external temperature 
from $0$ to $\infty$, the state with nonzero $\alpha$ becomes the partially mixed state even at the limit of $T \rightarrow \infty$. 
By making use of ${\mbox Tr} \rho_T^n$ the R\'{e}nyi and von Neumann entropies of $\rho_T$ are computed within ${\cal O} (\alpha)$ in section IV. The temperature-dependence
of the entropies is analyzed in this section. In section V a brief conclusion is given. In the appendix $A$, $B$, and $C$ we derive some integral formulas, which are 
used in section II.





\section{Thermal state}
Now, let us consider a simple harmonic oscillator in \textcolor[rgb]{1.00,0.00,0.00}{$(\widehat{P}, \widehat{Q})$-system}, whose Hamiltonian is 
\begin{equation}
\label{hamil-single}
\widehat{{\cal H}} = \frac{1}{2 m}\widehat{P}^2 + \frac{1}{2} m \omega^2 \widehat{Q}^2 = \frac{\widehat{p}^2}{2 m} + \frac{\alpha}{m} \widehat{p}^4 + \frac{1}{2} m \omega^2 \widehat{q}^2 + {\cal O} (\alpha^2).
\end{equation}
The Feynman propagator for this system was explicitly derived up to the first order of $\alpha$ in Ref. \cite{comment-1}, whose explicit expression is 
\begin{equation}
\label{GUP-kernel}
K[q_f, q_0: T] = \sqrt{\frac{m \omega}{2 \pi i \hbar \sin \omega T}} \left[ 1 + \alpha \textcolor[rgb]{0.00,0.00,1.00}{f(q_f, q_0: T)} + {\cal O} (\alpha^2) \right] e^{\frac{i}{\hbar} (S_0 + \alpha S_1)},
\end{equation}
where 
\begin{eqnarray}
\label{k-schrodinger-14}
&&S_0 = \frac{m \omega}{2 \sin \omega T} \left[ (q_0^2 + q_f^2) \cos \omega T - 2 q_0 q_f \right]                                                         \\     \nonumber
&&S_1 = - \frac{m^3 \omega^3}{32 \sin^4 \omega T}  \Bigg[ \left\{ 12 \omega T + 8 \sin 2 \omega T + \sin 4 \omega T \right\} (q_0^4 + q_f^4)               \\     \nonumber
&& \hspace{1.5cm} -4 \left\{ 12 \omega T \cos \omega T + 11 \sin \omega T + 3 \sin 3 \omega T \right\} q_0 q_f (q_0^2 + q_f^2)                                           \\     \nonumber
&& \hspace{4.5cm}   + 12 \left\{ 4 \omega T + 2 \omega T \cos 2\omega T + 5 \sin 2 \omega T \right\} q_0^2 q_f^2    \Bigg]                                               \\     \nonumber
&& \textcolor[rgb]{0.00,0.00,1.00}{ f(q_f, q_0: T)} = \frac{3 i \hbar m \omega}{8 \sin^2 \omega T}  \left( 2 \omega T + 5 \sin \omega T \cos \omega T + \omega T \cos 2 \omega T \right)                   \\     \nonumber
&& \hspace{2.8cm} - \frac{3 m^2 \omega^2}{8 \sin^3 \omega T} \Bigg[ 2 \omega T \left\{ 3 \cos \omega T (q_0^2 + q_f^2) -  2 (2 + \cos 2 \omega T )q_0 q_f  \right\}                     \\     \nonumber
&& \hspace{5.0cm} + 10 \sin \omega T (q_0^2 + q_f^2 - 2 q_0 q_f \cos \omega T) - 6 \sin^3 \omega T (q_0^2 + q_f^2)       \Bigg].
\end{eqnarray}

The Brownian or Euclidean propagator is defined as $G[q_f, q_0: \tau] = K[q_f, q_0: T= - i \tau]$. Then, the thermal state of 
the system is given by 
\begin{equation}
\label{thermal-1}
\rho_T [q_f, q_0: \beta] = \frac{1}{{\cal Z}} G[q_f, q_0: \hbar \beta]
\end{equation}
where $\beta = 1 / (k_B T)$. The parameters $k_B$ and $T$ are Boltzmann constant and external temperature, respectively. 
In Eq. (\ref{thermal-1})
${\cal Z}$ is a partition function defined as 
\begin{equation}
\label{partition-1}
{\cal Z} (\beta) \equiv \mbox{Tr} G = \int d q G[q, q: \hbar \beta].
\end{equation}
Then, the explicit expression of the thermal state becomes
\begin{eqnarray}
\label{thermal-2}
&&\rho_T[q_f, q_0: \beta]                                 \\    \nonumber
&&= \frac{1}{{\cal Z} (\beta)} \sqrt{\frac{m \omega}{2 \pi \hbar \sinh x}} \left[ 1 - 
\alpha \left( \textcolor[rgb]{0.00,0.00,1.00}{f_E (q_f, q_0: \beta)}  + \frac{1}{\hbar}  \textcolor[rgb]{0.00,0.00,1.00}{S_{1,E} (q_f, q_0: \beta)} \right) + {\cal O} (\alpha^2) \right]   \\   \nonumber
&&\hspace{4.0cm} \times\exp \left[-\frac{m \omega}{2 \hbar \sinh x} \left\{ (q_0^2 + q_f^2) \cosh x - 2 q_0 q_f \right\} \right]
\end{eqnarray}
where $x$ is a dimensionless parameter $x = \hbar \omega \beta$ and 
\begin{eqnarray}
\label{thermal-2-boso}
&& {\cal Z} (\beta) = \frac{1}{2 \sinh \frac{x}{2}} \left[ 1 - \frac{3 \alpha}{4} \left(\hbar m \omega x \right) \coth^2 \frac{x}{2} + {\cal O} (\alpha^2) \right]    \\    \nonumber
&& \textcolor[rgb]{0.00,0.00,1.00}{f_E (q_f, q_0: \beta)} = A_1 - A_2 (q_0^2 + q_f^2) + 2 A_3 q_0 q_f                           \\   \nonumber
&&\frac{1}{\hbar}  \textcolor[rgb]{0.00,0.00,1.00}{S_{1,E} (q_f, q_0: \beta)} = B_1 (q_0^4 + q_f^4) + B_2 (q_0^3 q_f + q_0 q_f^3) + B_3 q_0^2 q_f^2
\end{eqnarray}
with
\begin{eqnarray}
\label{thermal-2-boso-2}
&&A_1 = \frac{3 \hbar m \omega}{8 \sinh^2 x} \left[ 2 x + 5 \sinh x \cosh x + x \cosh 2 x \right]             \\    \nonumber
&&A_2 = \frac{3 m^2 \omega^2}{8 \sinh^3 x} \left[ 6 x \cosh x + 10 \sinh x + 6 \sinh^3 x \right]                         \\    \nonumber
&&A_3 = \frac{3 m^2 \omega^2}{8 \sinh^3 x} \left[2 x (2 + \cosh 2 x) + 10 \sinh x \cosh x \right]                        \\    \nonumber
&&B_1 = \frac{m^3 \omega^3}{32 \hbar \sinh^4 x} \left[ 12 x + 8 \sinh 2 x + \sinh 4 x \right]                                       \\    \nonumber
&&B_2 = -\frac{m^3 \omega^3}{8 \hbar \sinh^4 x} \left[ 12 x \cosh x + 11 \sinh x + 3 \sinh 3 x  \right]                  \\    \nonumber
&&B_3 = \frac{3 m^3 \omega^3}{8 \hbar \sinh^4 x} \left[4 x + 2 x \cosh 2 x + 5 \sinh 2 x \right].
\end{eqnarray}

In order to derive the  R\'{e}nyi and von Neumann entropies of $\rho_T[q_f, q_0: \beta]$ we may need to solve the eigenvalue equation 
\begin{equation}
\label{eigen-1}
\int d q_0 \rho_T[q_f, q_0: \beta] f_n (q_0: \beta) = \lambda_n f_n (q_f: \beta).
\end{equation}
Since $\mbox{Tr} \rho_T = 1$, the eigenvalue should obey $\sum_n \lambda_n = 1$. However, it is highly difficult to solve Eq. (\ref{eigen-1}), because 
the thermal state (\ref{thermal-2}) is not Gaussian when $\alpha \neq 0$. As we will show in the following, however, it is possible to derive the eigenvalue $\lambda_n$ exactly within the order of $\alpha$ without solving the eigenvalue 
equation (\ref{eigen-1}) explicitly.

\section{calculation of $\mbox{Tr} \rho_T^n$}
In this section we will compute $\mbox{Tr} \rho_T^n$ up to order of $\alpha$. 
One can show that $\mbox{Tr} \rho_T^n$ reduces to 
\begin{eqnarray}
\label{trace-1}
&&\mbox{Tr} \rho_T^n                                                                                                                                                            
\equiv \int dx_1 \cdots dx_n \rho_T[x_1, x_2: \beta] \rho_T[x_2, x_3: \beta] \cdots \rho_T[x_{n-1}, x_n: \beta] \rho_T[x_n, x_1: \beta]                                         \\    \nonumber
&&=\frac{1}{{\cal Z}^n} \left(\frac{m \omega}{2 \pi \hbar \sinh x} \right)^{n/2}                                                                                     \\    \nonumber
&& \times \int dx_1 \cdots dx_n 
\Bigg[ 1 - \alpha \bigg\{f_E (x_1, x_2: \beta) + \cdots +f_E(x_{n-1}, x_n: \beta) + f_E(x_n, x_1: \beta) \bigg\}                                                                  \\    \nonumber
&&\hspace{2.0cm} -\frac{\alpha}{\hbar} \bigg\{S_{1, E}(x_1, x_2: \beta) + \cdots + S_{1, E}(x_{n-1}, x_n: \beta) + S_{1, E} (x_n, x_1: \beta) \bigg\} + {\cal O} (\alpha^2)     \Bigg]         \\    \nonumber
&&\hspace{8.0cm}  \times \exp \left[- {\bm X} G_n {\bm X}^{\dagger} \right],
\end{eqnarray}
where ${\bm X} = (x_1, \cdots, x_n)$ and $G_n$ is a $n \times n$ matrix in a form
\begin{eqnarray}
\label{matrix-Gn}
G_n = \left(                  \begin{array}{cccccc}
                  2 a & \hspace{0.2cm}    -b    & \hspace{0.2cm}         & \hspace{0.2cm}         & \hspace{0.2cm}         & \hspace{0.2cm}  -b            \\
                  -b  & \hspace{0.2cm} 2 a      & \hspace{0.2cm} \bullet & \hspace{0.2cm}         & \hspace{0.2cm}         & \hspace{0.2cm}                \\
                      & \hspace{0.2cm} \bullet  & \hspace{0.2cm} \bullet & \hspace{0.2cm} \bullet & \hspace{0.2cm}         & \hspace{0.2cm}                \\
                      & \hspace{0.2cm}          & \hspace{0.2cm} \bullet & \hspace{0.2cm} \bullet & \hspace{0.2cm} \bullet & \hspace{0.2cm}                \\
                      & \hspace{0.2cm}          & \hspace{0.2cm}         & \hspace{0.2cm} \bullet & \hspace{0.2cm} \bullet & \hspace{0.2cm}   -b           \\
                   -b   & \hspace{0.2cm}          & \hspace{0.2cm}         & \hspace{0.2cm}         & \hspace{0.2cm}     -b  & \hspace{0.2cm}  2a 
                              \end{array}                        \right).
\end{eqnarray}
with
\begin{equation}
\label{Gn-boso-1}
a =\frac{m \omega \cosh x}{2 \hbar \sinh x}    \hspace{1.0cm}  b = \frac{m \omega}{2 \hbar \sinh x}.
\end{equation}
In Eq. (\ref{matrix-Gn}) matrix elements in empty space are zero. Inserting Eq. (\ref{thermal-2-boso}) into Eq. (\ref{trace-1}) $\mbox{Tr} \rho_T^n$ becomes
\begin{eqnarray}
\label{trace-2}
&&\mbox{Tr} \rho_T^n                                                                                                                                                            \\    \nonumber                                                                                                                                                          
&&=\frac{1}{{\cal Z}^n} \left(\frac{m \omega}{2 \pi \hbar \sinh x} \right)^{n/2}                                                                                     \\    \nonumber
&& \times \int dx_1 \cdots dx_n 
\Bigg[ 1 - \alpha \left\{n A_1 - 2 A_2 (x_1^2 + \cdots + x_n^2 ) + 2 A_3 (x_1 x_2 + \cdots + x_{n-1} x_n + x_n x_1) \right\}                                                                  \\    \nonumber
&&\hspace{3.0cm} -\alpha \bigg[2 B_1 (x_1^4 + \cdots + x_n^4)                                                                                                                               \\   \nonumber                                                                                                                            
&&\hspace{4.0cm}+ B_2 \left\{ x_1 x_2 (x_1^2 + x_2^2) + \cdots + x_{n-1} x_n (x_{n-1}^2 + x_n^2) + x_n x_1 (x_n^2 + x_1^2) \right\}                                                                       \\    \nonumber                                                                            
&& \hspace{4.0cm}                            + B_3 (x_1^2 x_2^2 + \cdots + x_{n-1}^2 x_n^2 + x_n^2 x_1^2 )  \bigg] + {\cal O} (\alpha^2)     \Bigg]                                                        \\    \nonumber       
&&\hspace{8.0cm}  \times \exp \left[- {\bm X} G_n {\bm X}^{\dagger} \right].
\end{eqnarray}
It is easy to show $\mbox{det} G_2 = 4 (a^2 - b^2)$, $\mbox{det} G_3 = 2 (a - b) (2 a + b)^2$, and $\mbox{det} G_4 = 16 a^2 (a^2 - b^2)$. Generalizing to $n$ one can show
\begin{equation}
\label{det-Gn}
\mbox{det} G_n = \frac{1}{2^n} \left[ \left( \sqrt{a + b} + \sqrt{a - b} \right)^n -  \left( \sqrt{a + b} - \sqrt{a - b} \right)^n \right]^2.
\end{equation}
Thus, it is possible to derive
\begin{eqnarray}
\label{int-formula-1}
&&\int dx_1 \cdots dx_n  \exp \left[- {\bm X} G_n {\bm X}^{\dagger} \right] = \frac{\pi^{n/2}}{\sqrt{\mbox{det} G_n}} \equiv g_n                                 \\     \nonumber
&&\int dx_1 \cdots dx_n (x_1^2 + \cdots + x_n^2) \exp \left[- {\bm X} G_n {\bm X}^{\dagger} \right]                                                              \\     \nonumber
&&= - \frac{1}{2} \frac{\partial g_n}{\partial a} = g_n \frac{n}{4 \sqrt{a^2 - b^2}} \frac{(\sqrt{a + b} + \sqrt{a - b})^n + (\sqrt{a + b} - \sqrt{a - b})^n}
                                                                                          {(\sqrt{a + b} + \sqrt{a - b})^n - (\sqrt{a + b} - \sqrt{a - b})^n}    \\     \nonumber
&&\int dx_1 \cdots dx_n (x_1 x_2 + \cdots + x_{n-1} x_n + x_n x_1)  \exp \left[- {\bm X} G_n {\bm X}^{\dagger} \right]                                           \\     \nonumber
&&= \frac{1}{2} \frac{\partial g_n}{\partial b} = g_n \frac{n b}{2 \sqrt{a^2 - b^2}} \frac{(\sqrt{a + b} + \sqrt{a - b})^{n-2} + (\sqrt{a + b} - \sqrt{a - b})^{n-2}}
                                                                                           {(\sqrt{a + b} + \sqrt{a - b})^n - (\sqrt{a + b} - \sqrt{a - b})^n}.
\end{eqnarray}
Furthermore, in appendix the following integral formula are derived:
\begin{subequations}
\label{int-formula-2}
\begin{equation}
\label{int-formula-2-1}
\int dx_1 \cdots dx_n (x_1^4 + \cdots + x_n^4)  \exp \left[- {\bm X} G_n {\bm X}^{\dagger} \right] = g_n \frac{3 n}{4} \frac{(\mbox{det} H_{n-1})^2}{(\det G_n)^2}    
\end{equation}
\begin{eqnarray}
\label{int-formula-2-2}
&&\int dx_1 \cdots dx_n (x_1^2 x_2^2 + \cdots + x_{n-1}^2 x_n^2 + x_n^2 x_1^2) \exp \left[- {\bm X} G_n {\bm X}^{\dagger} \right]                                       \\    \nonumber
&&= g_n \frac{n}{4 (\det G_n)^2} \Bigg[ 12 a^2 (\det H_{n-2})^2 - 12 a b^2 (\det H_{n-2}) (\det H_{n-3})                                                                 \\    \nonumber
&& \hspace{5.0cm}  + 3 b^4 (\det H_{n-3})^2 - 2 (\det H_{n-2}) (\det G_n) \bigg]                                                                                         
\end{eqnarray}
\begin{eqnarray}
\label{int-formula-2-3}
&&\int dx_1 \cdots dx_n \left[ x_1 x_2 (x_1^2 + x_2^2) + \cdots + x_{n-1} x_n (x_{n-1}^2 + x_n^2) + x_n x_1 (x_n^2 + x_1^2) \right]                                       \\    \nonumber
&& \hspace{8.0cm}  \times \exp \left[- {\bm X} G_n {\bm X}^{\dagger} \right]    \\    \nonumber
&&= g_n \frac{3 n}{2} \frac{ \left[ b^{n-1} + b (\det H_{n-2}) \right] \left[ 2 a (\det H_{n - 2}) - b^2 (\det H_{n - 3}) \right]}{(\det G_n)^2}
\end{eqnarray}
\end{subequations}
where $H_n$ is a $n \times n$ matrix defined in Eq. (\ref{matrix-Hn}) and its determinant is given by
\begin{eqnarray}
\label{det-Hn}
&&\det H_n = \frac{1}{2^n \sqrt{a^2 - b^2}} \Bigg[ a \left[ \left( \sqrt{a + b} + \sqrt{a - b} \right)^{2 n} - \left( \sqrt{a + b} - \sqrt{a - b} \right)^{2 n} \right]                 \\    \nonumber
&&      \hspace{4.0cm}                                      - b^2 \left[ \left( \sqrt{a + b} + \sqrt{a - b} \right)^{2 n - 2} - \left( \sqrt{a + b} - \sqrt{a - b} \right)^{2 n - 2} \right]  \Bigg].
\end{eqnarray}
Inserting Eqs. (\ref{thermal-2-boso-2}), (\ref{Gn-boso-1}), (\ref{int-formula-1}), and (\ref{int-formula-2}) into Eq. (\ref{trace-2}), one can derive
\begin{equation}
\label{trace-rho-n}
\mbox{Tr} \rho_T^n = \frac{2^{n-1} \sinh^n \frac{x}{2}} {\sinh \frac{n x}{2}}
\left[ 1 + \frac{3 \alpha n}{4} (\hbar m \omega x) \left( \coth^2 \frac{x}{2} - \coth^2 \frac{n x}{2} \right) + {\cal O} (\alpha^2) \right].
\end{equation}

From Eq. (\ref{trace-rho-n}) the purity function is given by 
\begin{equation}
\label{purity}
{\cal P} (T) \equiv \mbox {Tr} \rho_T^2 = \tanh \frac{x}{2} \left[ 1 + \frac{3 \alpha}{2} (\hbar m \omega x) \left( \coth^2 \frac{x}{2} - \coth^2 x \right) + {\cal O} (\alpha^2) \right].
\end{equation}
\begin{figure}[ht!]
\begin{center}
\includegraphics[height=6.0cm]{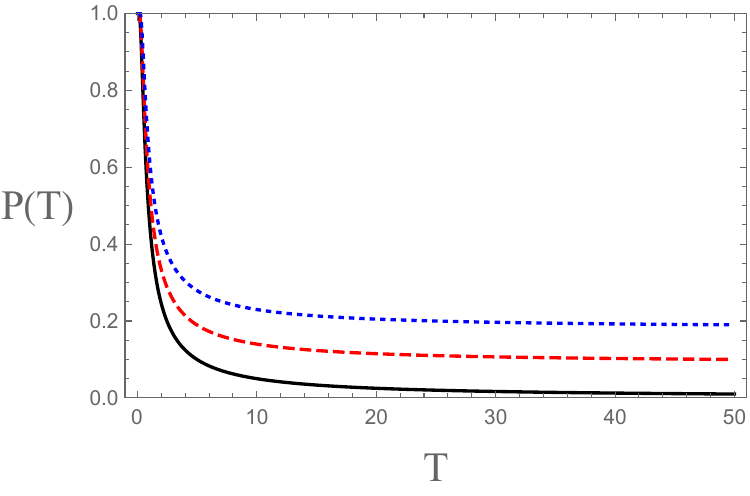} 

\caption[fig1]{(Color online) The external temperature dependence of the purity function when $\hbar = m = \omega = k_B = 1$. The black solid, red dashed, and blue dotted lines correspond to $\alpha = 0$, $\alpha = 0.04$, and $\alpha = 0.08$, respectively. As expected, the thermal state $\rho_T$ becomes pure and completely 
mixed state at $T = 0$ and $T = \infty$ when $\alpha = 0$. When, however, $\alpha \neq 0$, the state reduces to the partially mixed state even in the limit of $T \rightarrow \infty$.
 }
\end{center}
\end{figure}
The external temperature dependence of the purity function is plotted in Fig. 1 with varying the GUP parameter $\alpha$. For convenience we fix other constants as $\hbar = m = \omega = k_B = 1$.
In Fig. 1 the black solid, red dashed, and blue dotted lines correspond to $\alpha = 0$, $\alpha = 0.04$, and $\alpha = 0.08$, respectively. As expected, the thermal state $\rho_T$ becomes pure and completely 
mixed state at $T = 0$ and $T = \infty$ when $\alpha = 0$. When, however, $\alpha \neq 0$, the state reduces to the partially mixed state even in the limit of $T \rightarrow \infty$. This can be easily seen from 
$\lim_{T \rightarrow \infty} {\cal P}(T) = \frac{9 \alpha}{4}$. 

\section{R\'{e}nyi and von Neumann entropies}

\textcolor[rgb]{0.00,0.00,1.00}{In this section we will compute the R\'{e}nyi and von Neumann entropies of $\rho_T$, which are defined as 
\begin{equation}
\label{def-entropy}
S_{\gamma} = (1 - \gamma)^{-1} \ln \mbox{Tr} \rho_T^{\gamma}    \hspace{1.0cm} S_{von} = - \mbox{Tr} (\rho_T \ln \rho_T).
\end{equation}
It is worthwhile noting that $S_{von}$ can be obtained from $S_{\gamma}$ by taking a limit $\gamma \rightarrow 1$. }

\begin{figure}[ht!]
\begin{center}
\includegraphics[height=6.0cm]{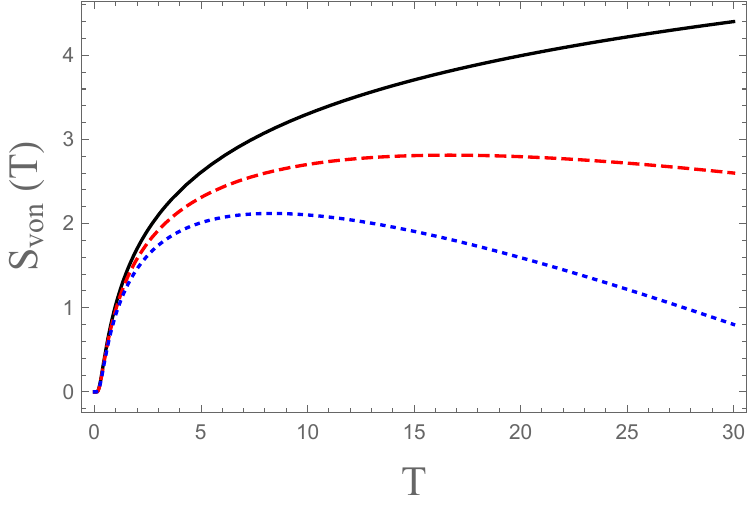} 

\caption[fig2]{(Color online) The external temperature-dependence of the von Neumann entropy when $\hbar = m = \omega = k_B = 1$. The black solid, red dashed, and blue dotted lines correspond to $\alpha = 0$, $\alpha = 0.01$, and $\alpha = 0.02$, respectively. 
When $\alpha = 0$, the von Neumann entropy monotonically increases with increasing temperature. The remarkable fact is that the von Neumann entropy for nonzero $\alpha$ exhibits a decreasing behavior at the high-temperature region. 
}
\end{center}
\end{figure}

Using Eq. (\ref{trace-rho-n}) it is possible to derive the eigenvalue $\lambda_n$ of the eigenvalue equation (\ref{eigen-1}) as following. 
Now, we take a trial solution of $\lambda_n$ in a form
\begin{equation}
\label{value-1}
\lambda_n (\beta) = (1 - e^{-x}) e^{-n x} \left[ 1 + \alpha (\hbar m \omega x) h_n (\beta) + {\cal O} (\alpha^2) \right].
\end{equation}
Since 
\begin{equation}
\label{value-2}
\sum_{n=0}^{\infty} \left( \lambda_n \right)^k = \mbox{Tr} \rho_T^k
\end{equation}
for all positive integer $k$, one can derive a condition of $h_n (\beta)$ as 
\begin{equation}
\label{value-3}
\sum_{n=0}^{\infty} e^{-n k x} h_n (\beta) = \frac{3}{4 \sinh^2 \frac{x}{2}} \frac{1}{1 - e^{-k x}} + \frac{3}{(1 - e^{- k x})^2} - \frac{3}{(1 - e^{- k x})^3}.
\end{equation}
Eq. (\ref{value-3}) can be solved as 
\begin{equation}
\label{value-4}
h_n (\beta) = 3 \left[ \frac{e^{- x}}{(1 - e^{- x})^2} - \frac{n (n + 1)}{2} + \zeta_n (\beta) \right],
\end{equation}
where $\zeta_n$ should satisfy $\sum_{n=0}^{\infty} e^{-n k x} \zeta_n (\beta) = 0$ for all positive integer $k$. Inserting Eq. (\ref{value-4}) into Eq. (\ref{value-1}), one can show
\begin{equation}
\label{value-5}
\lambda_n (\beta) = (1 - e^{-x}) e^{-n x} \left[ 1 + 3 \alpha (\hbar m \omega x) \left\{\frac{e^{- x}}{(1 - e^{- x})^2} - \frac{n (n + 1)}{2} + \zeta_n (\beta) \right\} + {\cal O} (\alpha^2) \right].
\end{equation}
It is straightforward to show $\sum_{n=0}^{\infty} \lambda_n = 1$, which is consistent with $\mbox{Tr} \rho_T = 1$. 

Now, we will show $\zeta_n (\beta) = 0$ as following. From Eq. (\ref{trace-rho-n}) the R\'{e}nyi entropy $S_{\gamma}$ with integer order can be explicitly derived.
If we take $\gamma \rightarrow 1$ limit in this expression, one can derive the von Neumann entropy in a form:
\begin{equation}
\label{von-1}
S_{von} = - \ln (1 - e^{-x}) +  \frac{ x e^{-x}}{1 - e^{-x}} 
- 3 \alpha (\hbar m \omega x^2) \frac{e^{-x} (1 + e^{-x})}{(1 - e^{-x})^3} + {\cal O} (\alpha^2).
\end{equation}
On the other hand the von Neumann entropy should be expressed as $-\sum_{n=0}^{\infty} \lambda_n \ln \lambda_n$. Comparing both expressions of the von Neumann entropy, one can conclude $\zeta_n (\beta) = 0$. 

\begin{center}
\begin{tabular}{c||cccccc} \hline \hline
$\alpha$ &    \hspace{0.2cm}   $0$  & \hspace{0.2cm} $0.01$  & \hspace{0.2cm} $0.02$  & \hspace{0.2cm} $0.03$  & \hspace{0.2cm} $0.04$  & \hspace{0.2cm} $0.05$    \\  \hline 
$T_*$    & \hspace{0.2cm} $\infty$  &  \hspace{0.2cm} $16.66$  & \hspace{0.2cm} $8.32$  & \hspace{0.2cm} $5.54$  & \hspace{0.2cm} $4.15$  & \hspace{0.2cm} $3.31$    \\  \hline  \hline
\end{tabular}

\vspace{0.1cm}
Table I:The $\alpha$-dependence of $T_*$ when $\hbar = m = \omega = k_B = 1$. 
\end{center}

The temperature dependence of the von Neumann entropy is plotted in Fig. 2 when $\hbar = m = \omega = k_B = 1$. The black solid, red dashed, and blue dotted lines correspond to $\alpha = 0$, $\alpha = 0.01$, and $\alpha = 0.02$, respectively.
Globally, the nonzero GUP parameter $\alpha$ reduces the von Neumann entropy in the whole range of $T$ compared to the case of $\alpha = 0$. This is due to the fact that the last term of Eq. (\ref{von-1}) has a minus sign. The remarkable fact is that while the von Neumann entropy with $\alpha = 0$ monotonically 
increases with increasing $T$, the von Neumann entropy with nonzero $\alpha$ has a maximum at finite temperature $T_*$. This temperature can be computed by solving 
$$(1 - e^{-\beta_* \omega})^2 - 3 \alpha (\hbar m \omega) \left[ (-2 + \beta_* \omega) + 4 \beta_* \omega e^{-\beta_* \omega} + (2 + \beta_* \omega) e^{-2 \beta_* \omega} \right] + {\cal O} (\alpha^2) = 0,$$
where \textcolor[rgb]{0.00,0.00,1.00}{$\beta_* = 1 / (k_B T_*)$.} Solving the above equation numerically, one can find the $\alpha$-dependence of temperature $T_*$. The $\alpha$-dependence of temperature $T_*$ is summarized in Table I when 
$\hbar = m = \omega = k_B = 1$. As Fig. 2 and Table I show, $T_*$ decreases with increasing $\alpha$. 
Mathematically, this extraordinary behavior of the von Neumann entropy with nonzero $\alpha$ can be understood as a competition of the last term in Eq. (\ref{von-1}) with the remaining terms. However, it is 
unclear, at least for us, to understand this behavior from a point of physics. 

\begin{figure}[ht!]
\begin{center}
\includegraphics[height=5.0cm]{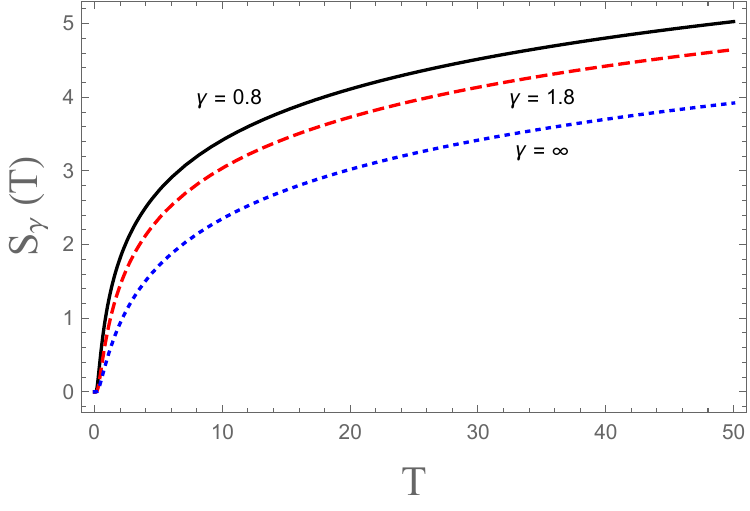} \hspace{1.0cm}
\includegraphics[height=5.0cm]{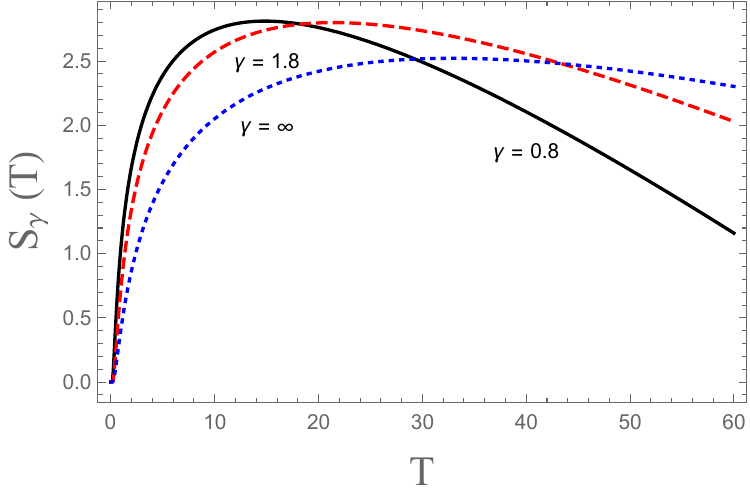}

\caption[fig4]{(Color online) The external temperature dependence of the R\'{e}nyi entropy  when $\hbar = m = \omega = k_B = 1$ when (a) $\alpha = 0$ and (b) $\alpha = 0.01$. In both figures the black solid, red dashed, and blue dotted lines correspond to $\gamma = 0.8$, $\gamma = 1.8$, and $\gamma = \infty$, respectively. 
When $\alpha = 0$, the R\'{e}nyi entropy decreases with increasing the order $\gamma$ in the whole range of $T$. This is well-known in the usual quantum mechanics.  The remarkable fact is that the R\'{e}nyi entropy with nonzero $\alpha$ exhibits a decreasing behavior at the high-temperature region. 
}
\end{center}
\end{figure}

The R\'{e}nyi entropy with arbitrary real order $\gamma$ is given by 
\begin{eqnarray}
\label{renyi-final}
S_{\gamma} &\equiv& \frac{1}{1 - \gamma} \ln \sum_{n=0}^{\infty} \left( \lambda_n \right)^{\gamma}                         \\    \nonumber
&=& \frac{1}{1 - \gamma} \Bigg[\gamma \ln (1 - e^{-x}) - \ln (1 - e^{-\gamma x})                     \\    \nonumber
&&       \hspace{2.0cm}    + 3 \alpha \gamma (\hbar m \omega x) \left\{ \frac{e^{-x}}{(1 - e^{-x})^2} - \frac{e^{-\gamma x}}{(1 - e^{-\gamma x})^2} \right\} + {\cal O} (\alpha^2) \Bigg].
\end{eqnarray}
In Fig. 3 the temperature dependence of the R\'{e}nyi entropy is plotted when $\alpha = 0$ (Fig. 3(a)) and $\alpha = 0.01$ (Fig. 3(b)) with choosing $\hbar = m = \omega = k_B = 1$. When $\alpha = 0$, the R\'{e}nyi entropies with various 
$\gamma$ exhibit increasing behavior with increasing temperature. With increasing the order $\gamma$ the entropy decreases gradually in the whole range of $T$ and eventually approaches to blue dotted line, which corresponds to $\gamma = \infty$. When $\alpha = 0.01$, however, 
the behavior of the R\'{e}nyi entropy is drastically changed at large temperature region. In this region the entropies exhibit decreasing behavior in temperature. Furthermore, with increasing the order $\gamma$ it increases and approaches to the
blue dotted line in this region. We do not exactly know how to interpret this extraordinary behavior physically.  

\section{Conclusion}
We compute in the paper the R\'{e}nyi and von Neumann entropies of the thermal state $\rho_T$ within the first order of $\alpha$. Both entropies with $\alpha = 0$ exhibit monotonically increasing behavior with increasing 
the external temperature. It is also found that the nonzero $\alpha$ decreases the entropies compared to the $\alpha = 0$ case.  Furthermore, there is a striking difference between $\alpha = 0$ and $\alpha \neq 0$. For the case of 
von Neumann entropy it has a maximum at finite temperature $T_*$ if $\alpha \neq 0$. In the region $T > T_*$ the von Neumann entropy exhibits a decreasing behavior with increasing the temperature. 
For the case of the R\'{e}nyi entropy $S_{\gamma} (T)$ it decreases with increasing the order $\gamma$ when $\alpha = 0$. If, however, $\alpha \neq 0$, the temperature dependence of the R\'{e}nyi entropy is drastically changed
at the large temperature region. In this region the R\'{e}nyi entropy exhibits a decreasing behavior with increasing the external temperature. The decreasing rate becomes larger with decreasing the order $\gamma$. As a result,
the R\'{e}nyi entropy becomes smaller with decreasing the order $\gamma$ at the large temperature region. 

It is of interest to extend our case to the one-dimensional two-coupled harmonic oscillator system, whose Hamiltonian is 
\begin{eqnarray}
\label{two-coupled-1}
\widehat{{\cal H}}_2 &=& \frac{1}{2 m} (\widehat{P}_1^2 + \widehat{P}_2^2 ) + \frac{1}{2} \left[ k_0 (\widehat{X}_1^2 + \widehat{X}_2^2) + J (\widehat{X}_1 - \widehat{X}_2)^2  \right]    \\     \nonumber  
&=& \widehat{h}_1 + \widehat{h}_2 + \frac{1}{2} J (\widehat{x}_1 - \widehat{x}_2)^2 + {\cal O} (\alpha^2),
\end{eqnarray}
where 
\begin{equation}
\label{two-coupled-2}
\widehat{h}_j = \frac{1}{2 m} \left( \widehat{p}_j^2 + 2 \alpha \widehat{p}_j^4 \right) + \frac{1}{2} k_0 \widehat{x}_j^2
\end{equation}
with $j = 1, 2$. It is possible to derive the vacuum and thermal states of this coupled system. It is of interest to compute the entanglement of formation\cite{benn96} of these states and compare them with 
the HUP cases\cite{park18,park19}. We guess the GUP parameter protects the entanglement degradation against the external temperature. However, it should be checked by explicit calculation. We hope to address this issue 
in the future.


\newpage 

\begin{appendix}{\centerline{\bf Appendix A: Derivation of Eq. (\ref{int-formula-2-1})}}

\setcounter{equation}{0}
\renewcommand{\theequation}{A.\arabic{equation}}
In this appendix we derive Eq. (\ref{int-formula-2-1}). First, let us consider a $n \times n$ matrix
\begin{eqnarray}
\label{matrix-tilde-Gn}
\tilde{G}_n = \left(                  \begin{array}{cccccc}
                  2 a' & \hspace{0.2cm}    -b    & \hspace{0.2cm}         & \hspace{0.2cm}         & \hspace{0.2cm}         & \hspace{0.2cm}  -b              \\
                  -b  & \hspace{0.2cm} 2 a      & \hspace{0.2cm} \bullet & \hspace{0.2cm}         & \hspace{0.2cm}         & \hspace{0.2cm}                \\
                      & \hspace{0.2cm} \bullet  & \hspace{0.2cm} \bullet & \hspace{0.2cm} \bullet & \hspace{0.2cm}         & \hspace{0.2cm}                \\
                      & \hspace{0.2cm}          & \hspace{0.2cm} \bullet & \hspace{0.2cm} \bullet & \hspace{0.2cm} \bullet & \hspace{0.2cm}                \\
                      & \hspace{0.2cm}          & \hspace{0.2cm}         & \hspace{0.2cm} \bullet & \hspace{0.2cm} \bullet & \hspace{0.2cm}   -b           \\
                   -b   & \hspace{0.2cm}          & \hspace{0.2cm}         & \hspace{0.2cm}         & \hspace{0.2cm}     -b  & \hspace{0.2cm}  2a 
                              \end{array}                        \right).
\end{eqnarray}
Then one can show
\begin{equation}
\label{app-a-1}
\det \tilde{G}_n = \det G_n + 2 (a' - a) \det H_{n-1}
\end{equation}
where $\det G_n$ is given by Eq. (\ref{det-Gn}) and $H_n$ is a following $n \times n$ tridiagonal matrix:
\begin{eqnarray}
\label{matrix-Hn}
H_n = \left(                  \begin{array}{cccccc}
                  2 a & \hspace{0.2cm}    -b    & \hspace{0.2cm}         & \hspace{0.2cm}         & \hspace{0.2cm}         & \hspace{0.2cm}                \\
                  -b  & \hspace{0.2cm} 2 a      & \hspace{0.2cm} \bullet & \hspace{0.2cm}         & \hspace{0.2cm}         & \hspace{0.2cm}                \\
                      & \hspace{0.2cm} \bullet  & \hspace{0.2cm} \bullet & \hspace{0.2cm} \bullet & \hspace{0.2cm}         & \hspace{0.2cm}                \\
                      & \hspace{0.2cm}          & \hspace{0.2cm} \bullet & \hspace{0.2cm} \bullet & \hspace{0.2cm} \bullet & \hspace{0.2cm}                \\
                      & \hspace{0.2cm}          & \hspace{0.2cm}         & \hspace{0.2cm} \bullet & \hspace{0.2cm} \bullet & \hspace{0.2cm}   -b           \\
                      & \hspace{0.2cm}          & \hspace{0.2cm}         & \hspace{0.2cm}         & \hspace{0.2cm}     -b  & \hspace{0.2cm}  2a 
                              \end{array}                        \right).
\end{eqnarray}
It is easy to show that $\det H_n$ satisfies the recursion relation
\begin{equation}
\label{app-a-2}
\det H_n = 2 a \det H_{n-1} - b^2 \det H_{n-2}
\end{equation}
with $\det H_0 = 1$ and $\det H_1 = 2 a$. Using Eq. (\ref{app-a-2}) it is possible to show
\begin{equation}
\label{app-a-3}
\det H_n = \frac{1}{\sqrt{a^2 - b^2}} \left[ a \sqrt{\det G_{2 n}} - \frac{b^2}{2} \sqrt{\det G_{2 n - 2}} \right]
\end{equation}
where $\det G_n$ is given in Eq. (\ref{det-Gn}). If Eq. (\ref{det-Gn}) is inserted into Eq. (\ref{app-a-3}), $\mbox{det} H_n$ reduces to 
Eq. (\ref{det-Hn}). 

Thus, it is easy to show 
\begin{eqnarray}
\label{app-a-4}
&&\int dx_1 \cdots dx_n x_1^4 \exp \left[- {\bm X} \tilde{G}_n {\bm X}^{\dagger} \right]                                   \\    \nonumber
&&= \frac{1}{4} \frac{\partial^2}{\partial a'^2}  \frac{\pi^{n/2}}{\sqrt{\det G_n + 2 (a' - a) \det H_{n-1}}}\Bigg|_{a' = a}    \\    \nonumber
&&= g_n \frac{3 \left(\det H_{n-1} \right)^2}{4 \left( \det G_n \right)^2}.
\end{eqnarray}

Next, by making use of the redefinition of the integral variables it is possible to show that 
$$ \int dx_1 \cdots dx_n x_j^4 \exp \left[- {\bm X} \tilde{G}_n {\bm X}^{\dagger} \right] $$
is independent of $j$. Thus, 
\begin{equation}
\label{app-a-5}
\int dx_1 \cdots dx_n \left(x_1^4 + \cdots + x_n^4 \right) \exp \left[- {\bm X} \tilde{G}_n {\bm X}^{\dagger} \right] 
= n \int dx_1 \cdots dx_n x_1^4 \exp \left[- {\bm X} \tilde{G}_n {\bm X}^{\dagger} \right],
\end{equation}
which exactly coincides with Eq. (\ref{int-formula-2-1}).

\end{appendix}

\newpage 

\begin{appendix}{\centerline{\bf Appendix B: Derivation of Eq. (\ref{int-formula-2-2})}}

\setcounter{equation}{0}
\renewcommand{\theequation}{B.\arabic{equation}}
In this appendix we derive Eq. (\ref{int-formula-2-2}). Let us consider a  $n \times n$ matrix
\begin{eqnarray}
\label{matrix-dasi-Gn}
G_n' = \left(                  \begin{array}{cccccc}
                  2 a_1 & \hspace{0.2cm}    -b    & \hspace{0.2cm}         & \hspace{0.2cm}         & \hspace{0.2cm}         & \hspace{0.2cm}  -b              \\
                  -b  & \hspace{0.2cm} 2 a_2     & \hspace{0.2cm} \bullet & \hspace{0.2cm}         & \hspace{0.2cm}         & \hspace{0.2cm}                \\
                      & \hspace{0.2cm} \bullet  & \hspace{0.2cm} 2 a & \hspace{0.2cm} \bullet & \hspace{0.2cm}         & \hspace{0.2cm}                \\
                      & \hspace{0.2cm}          & \hspace{0.2cm} \bullet & \hspace{0.2cm} \bullet & \hspace{0.2cm} \bullet & \hspace{0.2cm}                \\
                      & \hspace{0.2cm}          & \hspace{0.2cm}         & \hspace{0.2cm} \bullet & \hspace{0.2cm} \bullet & \hspace{0.2cm}   -b           \\
                   -b   & \hspace{0.2cm}          & \hspace{0.2cm}         & \hspace{0.2cm}         & \hspace{0.2cm}     -b  & \hspace{0.2cm}  2a 
                              \end{array}                        \right).
\end{eqnarray}
Then, one can show straightforwardly
\begin{equation}
\label{app-b-1}
\det G_n' = 4 (a_1 - a) (a_2 - a) \det H_{n-2} + 4 a (a_1 + a_2 - 2 a) \det H_{n-2} - 2 b^2 (a_1 + a_2 - 2 a) \det H_{n-3} + \det G_n.
\end{equation}
Thus, Eq. (\ref{int-formula-2-2}) is verified as following:
\begin{eqnarray}
\label{app-b-2}
&&\int dx_1 \cdots dx_n \left(x_1^2 x_2^2 + \cdots + x_{n-1}^2 x_n^2 + x_n^2 x_1^2 \right) \exp \left[- {\bm X} G_n {\bm X}^{\dagger} \right]           \\   \nonumber
&&= \frac{n}{4} \frac{\partial^2}{\partial a_1 \partial a_2} \frac{\pi^{n/2}}{\sqrt{\det G_n'}} \Bigg|_{a_1 = a_2 = a}                                          \\   \nonumber
&&= g_n \frac{n}{4 \left( \det G_n \right)^2} \Bigg[ 12 a^2 (\det H_{n-2})^2 - 12 a b^2 (\det H_{n-2}) (\det H_{n-3})                                           \\   \nonumber
&& \hspace{5.0cm} + 3 b^4 (\det H_{n-3})^2 - 2 (\det H_{n-2}) (\det G_n ) \Bigg],
\end{eqnarray}
which exactly coincides with Eq. (\ref{int-formula-2-2}).

\end{appendix}

\newpage 

\begin{appendix}{\centerline{\bf Appendix C: Derivation of Eq. (\ref{int-formula-2-3})}}

\setcounter{equation}{0}
\renewcommand{\theequation}{C.\arabic{equation}}
In this appendix we derive Eq. (\ref{int-formula-2-3}). Let us consider a  $n \times n$ matrix

\begin{eqnarray}
\label{matrix-twodasi-Gn}
G_n'' = \left(                  \begin{array}{cccccc}
                  2 a_1 & \hspace{0.2cm}    -b_1    & \hspace{0.2cm}         & \hspace{0.2cm}         & \hspace{0.2cm}         & \hspace{0.2cm}  -b              \\
                  -b_1  & \hspace{0.2cm} 2 a_1     & \hspace{0.2cm} -b & \hspace{0.2cm}         & \hspace{0.2cm}         & \hspace{0.2cm}                \\
                      & \hspace{0.2cm} -b  & \hspace{0.2cm} 2 a & \hspace{0.2cm} \bullet & \hspace{0.2cm}         & \hspace{0.2cm}                \\
                      & \hspace{0.2cm}          & \hspace{0.2cm} \bullet & \hspace{0.2cm} \bullet & \hspace{0.2cm} \bullet & \hspace{0.2cm}                \\
                      & \hspace{0.2cm}          & \hspace{0.2cm}         & \hspace{0.2cm} \bullet & \hspace{0.2cm} \bullet & \hspace{0.2cm}   -b           \\
                   -b   & \hspace{0.2cm}          & \hspace{0.2cm}         & \hspace{0.2cm}         & \hspace{0.2cm}     -b  & \hspace{0.2cm}  2a 
                              \end{array}                        \right).
\end{eqnarray}
Then, one can show straightforwardly
\begin{eqnarray}
\label{app-c-1}
&&\det G_n'' = \left[4 (a_1 - a) (a_1 + a) - (b_1 - b) (b_1 + b) \right] \det H_{n-2}                       \\    \nonumber
&& \hspace{2.0cm}- 4 b^2 (a_1 - a) \det H_{n-3} - 2 (b_1 - b) b^{n-1} + \det G_n.
\end{eqnarray}
Thus, Eq. (\ref{int-formula-2-3}) is proved as following:
\begin{eqnarray}
\label{app-c-2}
&&\int dx_1 \cdots dx_n \left\{x_1 x_2 (x_1^2 + x_2^2) + \cdots + x_{n-1} x_n (x_{n-1}^2 + x_n^2) + x_n x_1 (x_n^2 + x_1^2) \right\}                             \\   \nonumber   
&&  \hspace{9.0cm}       \times \exp \left[- {\bm X} G_n {\bm X}^{\dagger} \right]                                                         \\   \nonumber
&&= -\frac{n}{4} \frac{\partial^2}{\partial a_1 \partial b_1} \frac{\pi^{n/2}}{\sqrt{\det G_n''}} \Bigg|_{a_1 = a, b_1 = b}                                          \\   \nonumber
&&= g_n \frac{3 n}{2} \frac{ \left[ b^{n-1} + b (\det H_{n-2}) \right] \left[ 2 a (\det H_{n - 2}) - b^2 (\det H_{n - 3}) \right]}{(\det G_n)^2}
\end{eqnarray}
which exactly coincides with Eq. (\ref{int-formula-2-3}).

\end{appendix}

\end{document}